\documentclass[preprint,3p,helvetica,twocolumn]{elsarticle} 

\usepackage{graphicx,subfig,epsfig,wrapfig,xspace}
\usepackage{amssymb,amsthm} 
\usepackage[hyphens]{url} 
\usepackage{hyperref,breakurl}
\usepackage{color}

\newcommand{\snn}{\ensuremath{\sqrt{s_{{NN}}}}\xspace}
\newcommand{\pt}{\ensuremath{p_\mathrm{T}}\xspace}
\newcommand{\mt}{\ensuremath{m_\mathrm{T}}\xspace}
\newcommand{\ut}{\ensuremath{\langle u_\mathrm{T} \rangle}\xspace}
\newcommand{\etap}{\ensuremath{{\eta^\prime}}\xspace}

\newcommand{\Tcond}{\ensuremath{{T_\mathrm{cond}}}\xspace}
\newcommand{\Teff}{\ensuremath{{T_\mathrm{eff}}}\xspace}

\journal{Central European Journal of Physics} 

\begin{document}
\begin{frontmatter} 

\title{Effects of chain decays, radial flow and $U_{A}(1)$ restoration on the low-mass dilepton enhancement 
in \snn{}=200 GeV Au+Au reactions}

\author[auth1,auth2]{M\'arton Vargyas\fnref{mail1}}
\fntext[mail1]{vargyas@bnl.gov}
\address[auth1]{E\"otv\"os Lor\'and University, 
				\\ P\'azm\'any P\'eter s\'et\'any 1/a., 1117 Budapest, Hungary}

\author[auth2]{Tam\'as Cs\"org\H o\fnref{mail2}}
\fntext[mail2]{csorgo.tamas@wigner.mta.hu}
\address[auth2]{Wigner Research Centre for Physics of the Hungarian Academy of Sciences, 
				\\Konkoly-Thege Mikl\'os \'ut 29-33., 1121 Budapest, Hungary}

\author[auth2,auth3]{R\'obert V\'ertesi \fnref{mail3}}
\fntext[mail3]{robert.vertesi@ujf.cas.cz}
\address[auth3]{Nuclear Physics Institute of The Academy of Sciences of the Czech Republic, \\ 
					25068 \v{R}e\v{z}, Czech Republic}

\begin{abstract} 
In \snn{}=200 GeV Au+Au collisions PHENIX reported a significant enhancement in the low-mass region (0.1$<${}$m_{\mathrm{ee}}${}$<$0.7 GeV) of the dielectron spectrum, which is still not fully understood.
Several theoretical works and an indirect measurement suggest, due to the possible restoration of the $U_A(1)$ part of the chiral symmetry in a hot and dense medium, that the mass of the \etap meson may substantially decrease. 
This work reports on a statistically acceptable description of the PHENIX low-mass dilepton enhancement using a radial flow dominated meson spectra, chain decays of long-lived resonances and an in-medium \etap mass modification.
\end{abstract} 


\end{frontmatter}

\section{Motivation}
In comparison with the expected hadronic decays, PHENIX measured a significant excess of $\textrm{e}^{+}\textrm{e}^{-}$ pairs in the invariant mass range of 0.1--0.7 GeV in Au+Au collisions~\cite{PPG088}, while no such excess is present in p+p collisions~\cite{Adamczyk:2012yb}. 
Preliminary results from STAR also show a substantial, although smaller effect in a different geometrical acceptance~\cite{Geurts:2012rv}.
This low-mass dilepton enhancement is the subject of detailed theoretical investigations, and, according to a recent conference report~\cite{Tserruya:2012jb}, the enhancement is not yet fully understood.

Lattice results~\cite{Fodor:2009ax} suggest that the transition to chirally symmetric and deconfined matter is a smooth, cross-over type transition, where the chiral symmetry restoration as measured by chiral susceptibility happens at a significantly lower temperature than the confinement of quarks, as measured by the Polyakov loop. The continuum extrapolated value for the transition temperature of the renormalized Polyakov loop is 25$\pm$4 MeV higher than that of the chiral susceptibility. This result suggests an interesting possibility of chiral hadron dynamics in the 150--175 MeV temperature range.
Chiral symmetry restoration may change various properties of hadrons, such as their mass, width, branching ratios and decays.
The PHENIX collaboration was the first to publish the direct photon spectrum in \snn{}=200 GeV Au+Au collisions~\cite{Adare:2008ab}, where the average temperature of the photon emitted region was found to be 220$\pm$19{\it(stat)}$\pm$19{\it(syst)} MeV,
implying that the initial temperature was at least about 300 MeV, significantly larger than both the Hagedorn temperature (the upper limit of the existence of hadrons, $\sim$170 MeV) and the kinetic freeze-out temperature (where hadronic interactions cease, in the order of $m_\pi${}$\sim$140 MeV.)

The in-medium drop of the mass of the \etap meson was anticipated a long time ago~\cite{Kunihiro,Kapusta,Huang:1995fc}.
The $\etap$ meson is different from the other pseudoscalar mesons in the sense that, due to the axial anomaly, it has a very large mass ($m_{\etap} = 958$ MeV) which prevents it from acting as the 9\textsuperscript{th} pseudo-Goldstone boson. However, in a hot and dense hadronic medium where the chiral symmetry is partially restored, the mass of the \etap can be reduced substantially, to the proximity of its quark model value ($\sim$500 MeV), to the mass-range of the other pseudoscalar mesons.
The first indirect observation of in-medium chiral dynamics, based on Bose-Einstein correlation (BEC) measurements and on the analysis of the $\etap${}$\rightarrow${}$\pi^+\pi^- \eta${}$\rightarrow${}$\pi^+\pi^- \pi^+\pi^-(\pi^0/\gamma)$   decay chain, suggested a significant and an at least 200 MeV in-medium mass drop of the \etap mesons in these reactions ~\cite{Vertesi:2009io,Vertesi:PRC,Vertesi:2009ca}.
These results triggered renewed interest in theoretical investigations of $U_A(1)$ symmetry restoration: the Witten-Veneziano formula has been generalized to finite temperature to check if the indirectly measured in-medium \etap masses are in the theoretically acceptable range~\cite{Benic:2011fv,Benic:2012eu,Kwon:2012vb}, lattice QCD calculations were performed to study the temperature dependence of the $U_A(1)$ symmetry restoration~\cite{:2012ja}. In cold nuclear matter, the nuclear bound states and partial restoration of the \etap mass was investigated in Ref.~\cite{Jido:2011pq}, to mention just some of the most inspiring new results. 

The indirect observation of in-medium \etap mass modification in Refs.~\cite{Vertesi:2009io,Vertesi:PRC,Vertesi:2009ca} is based on a hadronic decay channel. In this work, we cross-check this effect in the dilepton spectrum, utilizing the $\etap${}$\rightarrow${}$ \mathrm{e}^{+}\mathrm{e}^{-}\gamma$ (Dalitz) decay channel. This channel contributes to the low mass region (below $\sim$1 GeV), where the observed excess is present~\cite{PPG088}.
We also show that radial flow effects play an important role in understanding the measured dilepton spectrum.
Since the new STAR and PHENIX data~\cite{Geurts:2012rv,Tserruya:2012jb} are preliminary, we restrict our studies to the published PHENIX data~\cite{PPG088}.

We have to keep in mind that other meson properties may also change in the hot and dense hadronic medium. 
The disappearance of the $U_A(1)$ anomaly predominantly affects the \etap meson, and therefore no significant mass drop is anticipated for the other mesons due to the restoration of this symmetry. However, strong interactions on the hot and dense, hadronic matter may lead to additional mass modifications as well as collisional broadening of resonance decay widths, see e.g\ Ref.~\cite{KB} for a review.
This is usually formalized through the time evolution of spectral functions as given by the Kadanoff-Baym equations~\cite{KB}. From Eq.~(157) of that paper, the collisional broadening of the \etap can be estimated to be approximately half of that of the $\phi$ meson, well below the 50 MeV binning of the PHENIX dilepton spectrum of Ref.~\cite{PPG088}. Thus in the present paper we do not consider these effects. This is in contrast to those cases where the interactions among the different mesons play a non-negligible role, such as a possible collisional broadening of the $\rho$ meson, where even a relatively small increase of the 150~MeV vacuum width is experimentally accessible.

A recent measurement of the TAPS collaboration~\cite{Nanova:2012vw} reported such broadening of the \etap in cold nuclear matter to $\Gamma_\etap^*${}$\sim$15--25 MeV from the vacuum value of $\Gamma_\etap${}$\sim$0.2 MeV. Results from ANKE \cite{AN1,AN2} also point towards an enormous broadening of mesons in cold nuclear matter. The broadening of the $\phi$ meson width from 4 MeV to 35-50 MeV \cite{AN1} is approximately two times that of the broadening observed for the \etap, and it is comparable to the 50 MeV bin size of the PHENIX measurement. Both the $\etap$ and $\eta$ resonances are, however, very narrow ($\Gamma_{\eta'}$=1.30$\pm$0.07 keV and $\Gamma_\eta$=204$\pm$15 keV).
The sensitivity of the present analysis to broadening effects is predominantly limited by the statistics in the range of the low mass dilepton enhancement. Although the temperature dependence of the broadening effects is not clear, the fact that our simulations reproduce the position, the shape and also the transverse momentum dependence of the $\phi$ meson within 1--2 standard deviations without further tuning, gives us confidence that a possible broadening of the 20 times narrower \etap is not observable within the current bin sizes. More detailed experimental data, however, may allow for a direct observation of the possible broadening of the width of \etap in hot hadronic matter.

The NA60 Collaboration reported the first measurement of the broadening of the $\rho$ spectral function~\cite{Arnaldi:2006jq}, based on a low-mass dilepton measurement  in a similar kinematic range as the PHENIX low mass dilepton enhancement, in $E_{Lab} =158$ AGeV In-In collisions at the CERN SPS~\cite{Arnaldi:2006jq}. The associated  spectral function of the $\rho$ meson indicates significant broadening, but the position of the in-medium $\rho$ mass peak was found to be in agreement with the vacuum value within experimental errors~\cite{Arnaldi:2006jq}. The same mechanism, however, is not able to fully account for the PHENIX low-mass dilepton enhancement in \snn{}=200 GeV Au+Au reactions~\cite{Tserruya:2012jb,KB}. 
Also note that, contrary to RHIC collisions, a previous analysis of NA44 S+Pb data in the BEC channel showed no significant \etap mass drop at \snn{}=19.4 GeV~\cite{Kharzeev}. This indicates an expressive change of chiral dynamics between the CERN SPS top energy of \snn{}=19.4 GeV and the RHIC energies of \snn{}=60, 130 and 200 GeV (see Ref.~\cite{Kunihiro} for details on this point).
The method we utilize here is fully consistent with the methods applied in Refs.~\cite{Vertesi:2009io,Vertesi:PRC}, which is an indirect \etap mass modification measurement using Bose-Einstein correlations of pions from \etap chain decays. As far as we know (also see Ref.~\cite{Tserruya:2012jb}), this is the first statistically acceptable description of the low-mass dilepton enhancement in 200 GeV Au+Au collisions at RHIC, as we shall demonstrate below. Still, there are several possibilities for further improvements. Ideally all the in-medium effects should be considered simultaneously. In this analysis, however, we restrict ourselves in exploring the complex phenomenon of chiral symmetry restoration only to the mass modification of the \etap meson, which is expected to be the most prominent effect, especially at the lower temperatures of the hadronic phase. 

\section{The $\etap$ meson in a hot, dense medium}
\subsection{Effect of mass-modification}
As a result of a possible chiral symmetry restoration the mass of the $\etap$ meson can change significantly. This mass modification, however, cannot be observed directly due to the small inelastic and annihilation cross-section ($\sigma_0^\etap${}$\sim$2.6 mb) and the large lifetime ($>$10 fm/$c$) of the \etap in the medium\footnote{%
	We denote the properties of the in-medium $\etap$ with an asterisk, while quantities without further notation correspond to the free $\etap$ throughout this paper.}. Similarly to our previous investigations in~\cite{Vertesi:2009io,Vertesi:PRC,Vertesi:2009ca} we thus neglect the decays of $\eta'^*$ in the medium, but consider how the in-medium $\eta'^*$ enhancement can be observed after the medium dissolves and these $\eta'^*$ mesons become on-shell and decay in the vacuum. This enhancement can be described with a modified Hagedorn-formula~\cite{Kharzeev,Csorgo:1995bi}:
	
\begin{equation}
f_{\etap} = \left(\frac{m_{\etap}^*}{m_{\etap}}\right)^\alpha e^{\frac{m_{\etap}-m_{\etap}^*}{\Tcond}}, \label{e: Hagedorn}
\end{equation}
where $m_{\etap}$ is the vacuum mass-, $m_{\etap}^*$ is the in-medium mass of the $\etap$ meson, while \Tcond is the characteristic inverse slope of the \etap mesons within the formed medium.
 
Note that many of the arguments presented in this section resemble earlier ones that considered a possible low transverse momentum enhancement of photons and pions, K$^{-}$ mesons and antiprotons~\cite{S1,S2,K1}\footnote{%
	Eg.\ ref.~\cite{S1} considered a hypotethical attractive mean field potential acting as a
	trap for low transverse momentum kaons to explain preliminary data on low \pt enhancement
	of K$^{-}$. Our arguments specifically relate to the connection of the \etap meson to the
	$U_A(1)$ anomaly and the resulting well known and large mass difference between the 
	$\eta$ and \etap mesons, as emphasized in the context of heavy ion phenomenology first in
	refs.~\cite{Kunihiro,Kapusta,Huang:1995fc}.
}.
In order to explore the role of $\etap$ meson enhancement in the $\sqrt{s_{NN}} = 200$ GeV Au+Au collision data of PHENIX~\cite{PPG088}, extended EXODUS simulations were elaborated to generate different dilepton spectra corresponding to physically reasonable $\etap$ properties. 
\begin{figure}[h]
 \centering
  {\includegraphics[width=0.45\textwidth]{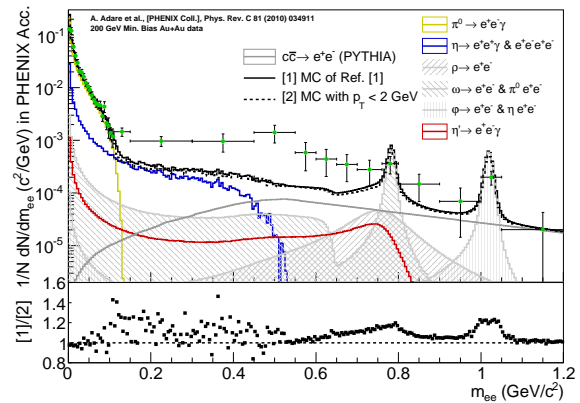} }
  \caption{PHENIX dilepton cocktail of Ref.~\cite{PPG088} in PHENIX acceptance, reproduced by our simulations (solid black line), compared to the same cocktail with a \pt$<$ GeV cut applied for each resonance (dashed black line). Points are PHENIX data of Ref.~\cite{PPG088}, colored lines are different cocktail components with the \pt cut applied. The bottom panel shows the ratio of the cocktails with and without the \pt cut.}
  \vspace{-10pt}
  \label{f: PPG088}
\end{figure}

\subsection{The shape of the $\etap$ spectrum}
According to the conservation of energy, those $\etap$ mesons that can escape from the hot and dense hadronic medium, while regaining their mass, must lose momentum. As we compare our simulations to data in the PHENIX acceptance ($|\eta| < 0.35$), we neglect the longitudinal component of the momentum for the purpose of this work.
Thus the conservation of energy requires $\etap$ mesons to fulfill the 
\begin{equation} \label{e: cons}
{m_{\etap}^*}^2 + {\pt^*}^2 = m_{\etap}^2 + \pt^2
\end{equation}
relation. 
This condition results in two different components of the $\etap$ spectrum. Both components realize an effective $|0\rangle_*$+$|\etap\rangle_*${}$\rightarrow${}$|0\rangle$+$|\etap\rangle$ transition, but in a different way.

Some of the $\etap$ meson's transverse momenta is large enough to let them escape from the hot and dense $U_A(1)$ restored medium to the asymptotic vacuum by decreasing their momenta and increasing their mass. They yield the first component of the \etap spectrum, which is dominant in the \pt{}$>$\Teff=$T_0 + m \ut^2$ region and follows a nearly exponential spectrum similar to other hadrons, and can be described with the same hydrodynamical model.
Those $\etap$ mesons, however, that have low transverse momentum\footnote{%
	Low transverse momentum means $\etap$ mesons with 
	$\pt^* < \sqrt{{m_{\etap}}^2 - {m_{\etap}^*}^2}$ 
	as a result of conservation of energy (Eq.~\ref{e: cons})
} in the chirally modified vacuum, cannot come on mass shell thus they are trapped in that modified vacuum. They are released and come on-shell only after the dissociation of the modified vacuum state to the asymptotic vacuum, thus they are emitted with a small \pt, characteristic to the decaying condensate. This second component of the \etap spectrum may be dominant at low-\pt if there is a significant in-medium \etap mass decrease. The trapped $\etap$ mesons are assumed to follow a Maxwell--Boltzmann-like distribution,
\begin{equation} \label{e: etap_low_pT}
f(\pt) = \frac{1}{2\pi m_{\etap}B^{-1}} e^{- \frac{p_{T}^2}{2m_{\etap}B^{-1}} } \ ,
\end{equation}
where $B^{-1}$ is the slope parameter of the $\etap$ condensate. 

The final $\etap$ spectrum is the sum of Eq.~(\ref{e: etap_low_pT}) and the hydrodynamical spectrum, which results in a double exponential spectrum as used in Refs.~\cite{Vertesi:2009io,Vertesi:PRC} and also in the current analysis.

\section{Radial flow for dilepton enhancement} \label{sec: radial flow}
Another idea which might bring us closer to understand the dilepton excess is to introduce a new \pt-distribution with radial flow for the mesons. The radial flow leads to larger effective slope parameters for heavier resonances in the soft, 0$<$\pt{}$<$2 GeV part of the spectrum. In contrast, PHENIX used a background where resonance spectra were extrapolated to this region using \mt-scaling. Radial flow breaks the \mt-scaling of single particle spectra. This effect, as far as we know, was not included in the original PHENIX analysis~\cite{PPG088}. Radial flow is valid in the low \pt range, where it describes particle spectra well. This region is also the most important one for the dilepton spectrum. In Fig.~\ref{f: PPG088} the original PHENIX result~\cite{PPG088} and a resonance \pt{}$<$2 GeV) cut result are plotted, confirming the dominance of the low-\pt part of the spectra.
Thus examining the effect of the radial flow seems to be inevitable, as it might be responsible for certain parts of the excess.

Previous examinations suggested that a spectrum motivated by perfect hydrodynamics can give a better estimate of the low-\pt regime for all mesons. We made use of the following simplified hydrodinamical formula, which was first derived for non-central heavy-ion collisions as an exact solution of hydrodynamics \cite{Csorgo:2001xm}, but it is also applied in phenomenological models~\cite{Csorgo:1995bi} and in data analysis~\cite{Adler:2003cb}:

\begin{equation} \label{e: hydro}
E\frac{d^3N}{d p^3} = A  \left(\frac{\mt}{m}\right)^\alpha  e^{-\frac{\mt-m}{T_0 + m \ut^2}}
\end{equation}

\begin{figure}[t]
 \begin{center}
 \includegraphics[width=0.45\textwidth]{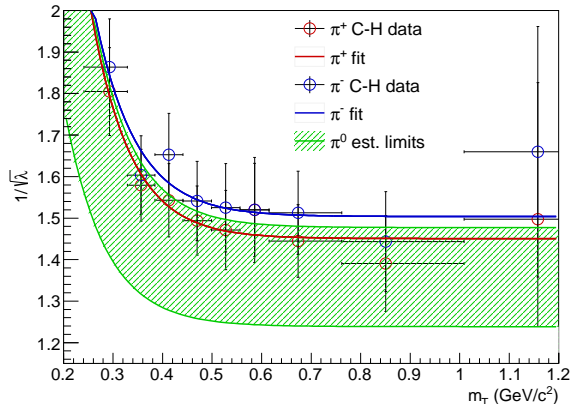} 
 \caption{Core-Halo correction parameter, measuring the ratio of particles in the core compared to the total number of particles (including halo). The upper and the lower boundaries of the $\pi^0$ correction band are the average of charged pion corrections and half of the average charged pion corrections respectively.} \label{f: C-H}
  \vspace{-20pt}
 \end{center}      
\end{figure}

\begin{figure}[t]
 \begin{center}
 \includegraphics[width=0.45\textwidth]{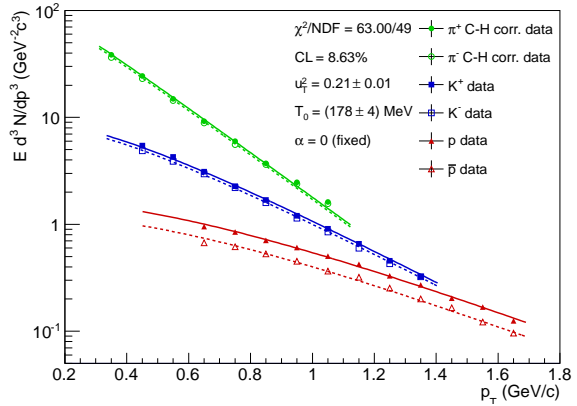} 
 \caption{Simultaneous fit of Eq.~(\ref{e: hydro}) for  $\pi^{\pm}$, $K^{\pm}$, $p$ and $\bar{p}$ data of Ref.~\cite{Adler:2003cb}, with the following constraint for the fit ranges in (\mt{}$-${}$m$): 0.2--1.0 GeV/$c^2$ for $\pi^{\pm}$ and 0.1--1.0 GeV/$c^2$ for $K^{\pm}$, $p$ and $\bar{p}$. The charged $\pi$ spectra are Core-Halo corrected.%
 \label{f: simultan fit}}
  \vspace{-10pt}
 \end{center}      
\end{figure}

Hydrodynamical models usually describe the primordially created particles, but not their decay products. This effect is important in case of the $\pi^{0}$ meson, which can be a daughter of many other hadrons. Thus we corrected the measured pion spectrum using the Core-Halo (C-H) picture~\cite{Csorgo:1995bi,Lednicky:1979ig,Bolz:1992ye,Csorgo:1994in}. Primordial pions (together with decay products of rapidly decaying resonances that yield thermalized pions) are in the core, the hydrodynamically evolving medium, while decay product pions of long lived resonances form the halo farther away from the collision point. Thus if we divide the measured $\pi$ spectrum with
\begin{equation}
\frac{1}{\sqrt{\lambda}} = \frac{N_{core}+N_{halo}}{N_{core}},
\end{equation}
we get an estimate for the number of core pions. An upper boundary on the corrections for neutral pions was obtained by averaging the $\pi^+$ and $\pi^-$ C-H corrections~\cite{Adler:2004rq}, $\lambda_H$=$(\lambda_{\pi^+}+\lambda_{\pi^-})/2$, while the lower boundary was $\lambda_L$=$\frac{\sqrt{\lambda_H}}{2}(1+\sqrt{\lambda_H})$, based on the consideration that long lived resonances do not produce more than twice as much charge as neutral pions. The results are shown in Fig.\ \ref{f: C-H}. Our $\pi^\pm$ spectra are  C-H corrected, even before the fit of Eq.~(\ref{e: hydro}) is applied.
The hydrodinamical formula of Eq.~(\ref{e: hydro}) was simultaneously fitted to available $\pi^{\pm}$, $K^{\pm}$, $p$ and $\bar{p}$ minimum bias data of Ref.~\cite{Adler:2003cb}, with the following constraint for the fit ranges in (\mt{}$-${}$m$): 0.2--1.0 GeV/$c^2$ for $\pi^{\pm}$ and 0.1--1.0 GeV/$c^2$ for $K^{\pm}$, $p$ and $\bar{p}$ (also from Ref.~\cite{Adler:2003cb}). The fit describes the spectra, as indicated in
Fig.~\ref{f: simultan fit}.
The resulting fit parameters are $\ut^2$=0.21$\pm$0.01 and $T_0$= 178$\pm$4 MeV, while the $\alpha$ parameter was within errors consistent with 0, hence it was fixed to 0. The fit quality is $\chi^2/$NDF=63/49, CL=8.6\%.

\section{Dilepton cocktail simulations} \label{sec:simulations}
\subsection{Radial flow and chain decays} \label{ssec: radial flow}
Every meson was generated with the hydrodynamical spectrum of Eq.~(\ref{e: hydro}), except for the $\etap$. Their normalization factors, however, could not be determined in every case, due to the lack of data in the valid range for hydrodynamics (\pt{}$\lesssim$2~GeV). Moreover, these inclusive spectra contain feed-down contributions from resonances besides directly produced (core) particles. Therefore in this work we used thermal model predictions of Ref.~\cite{Kaneta} for the core hadrons, and we simulated their decay chains using Monte-Carlo methods.

In the PHENIX and STAR cocktails~\cite{PPG088,Adamczyk:2012yb,Geurts:2012rv,Tserruya:2012jb} the \etap production is assumed to be small, thus the \etap chain decay effects, e.g.\ $\eta'${}$\rightarrow${}$\eta+\pi^0+\pi^0${}$\rightarrow${}$ (\textrm{e}^+\textrm{e}^-\gamma)+(\textrm{e}^+\textrm{e}^-\gamma)+(\textrm{e}^+\textrm{e}^-\gamma)$ are also estimated to be negligible. In our scenario, however, the \etap production may be increased by a factor as large as 50 \cite{Vertesi:2009io,Vertesi:PRC,Vertesi:2009ca}, hence feed-down from \etap to $\eta$ and to other resonances that decay subsequently to dileptons have to be considered. In our work all direct and cascade decay channels of $\pi^{0}$, $\omega$, $\rho$, $\eta$, \etap and $\phi$ that produce lepton pairs were individually simulated. We used the EXODUS Monte-Carlo-based event and decay generator. EXODUS is a fast MC event generator tuned to the PHENIX acceptance, which generates hadron distributions according to a previously specified \pt spectrum and 
computes their two- and three-body decays. EXODUS was also used by PHENIX to evaluate their background dilepton cocktail, as detailed in Ref.~\cite{PPG088}.

\begin{figure}[h]
 \begin{center} 
  \subfloat{ \includegraphics[width=0.45\textwidth]{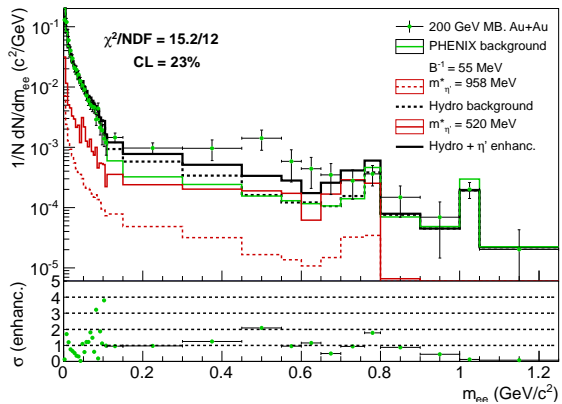} } 
\caption{Dilepton spectra in PHENIX acceptance with the perfect hydrodynamics motivated spectrum of Eq.~(\ref{e: hydro}) including chain decays.
The dashed red line is the contribution of the unmodified \etap, the solid one is the contribution from the best fit, $m_{\etap}^*$=520 MeV. Dashed and solid black lines are the corresponding total cocktails. Points are PHENIX data, the green line is the reproduced cocktail~\cite{PPG088}.}\label{f: sim_result} 
  \vspace{-20pt}
 \end{center}     
\end{figure}

The simulation results, with the input spectra of Eq.~(\ref{e: hydro}) and including chain decays, are shown with dashed lines in Fig.\ \ref{f: sim_result}. The agreement with data is improved from $\chi^2/$NDF=50.3/12, CL=0.01\% (cocktail of Ref.~\cite{PPG088}) to $\chi^2/$NDF=24.1/12, CL=1.9\%, both calculated for the 0.12--1.2 GeV mass range. This implies that radial flow effects on the hadronic background cocktail alone significantly improve the agreement of simulations with PHENIX data.

\subsection{Including \etap mass modification} \label{ssec: etap_enh}
\begin{figure}[t]
 \begin{center} 
  \vspace{-15pt}
   \includegraphics[width=0.48\textwidth]{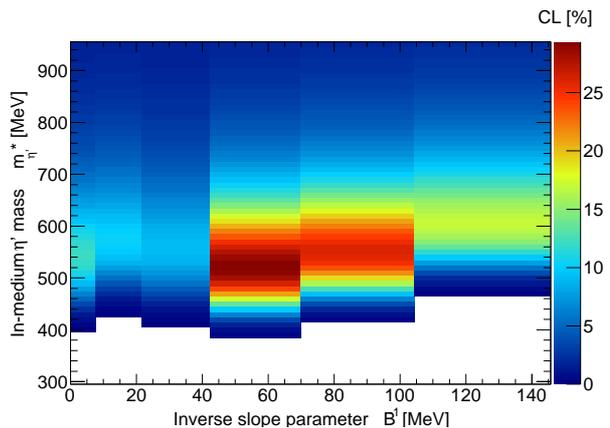} 
\caption{Preliminary confidence level map of the two $\etap$ spectrum parameter $B^{-1}$ and $m_{\etap}^*$. Best fit corresponds to $m_{\etap}^*${}$\approx$520 MeV and $B^{-1}${}$\approx$55 MeV. The $m_{\etap}^*${}$<$380 MeV region (white) corresponds to CL$<$0.1\% and can be excluded, but no upper limit can be given.
} \label{f: CL_map} 
  \vspace{-20pt}
 \end{center}     
\end{figure}

In the previous subsection we evaluated the new, hydrodynamical background with chain decays. In this section we combine it with the possible effect of the $\etap$ mass modification.
We determined the most probable value of the number of produced \etap mesons and the corresponding enhancement factor $f_\etap$ by comparing the results to the measured dilepton spectrum. We concluded the in-medium mass of $\etap$ from $f_\etap$ by substituting it into Eq.~(\ref{e: Hagedorn}), using \Tcond=170 MeV~\cite{Vertesi:PRC}.
The \etap spectrum is technically determined by two parameters: the inverse slope parameter, $B^{-1}$ (see Eq.~(\ref{e: etap_low_pT})) and the in-medium \etap mass, $m_{\etap}^*$. 
We scanned through the physically meaningful range of these two parameters in order to find their most probable values. In order to reduce the simulation time, we introduced a simplification to this step: we handled $m_{\etap}^*$ as a parameter which governs only the overall size of the \etap excess (see Eq.~(\ref{e: Hagedorn})), while $B^{-1}$ is responsible for the shape of the spectrum. In fact, the slope of the longer component of the excess is also influenced by the in-medium mass, but the effect was found to be small when compared to the \pt-integrated dilepton spectrum.
Fig.\ \ref{f: CL_map} shows the confidence level map of these two parameters calculated for the 0.12--1.2 GeV region.
The confidence level map in Fig.\ \ref{f: CL_map}  indicates the first results of the scan of the in-medium \etap mass and the slope parameter $B^{-1}$. 
This map indicates that very low $\eta_*'$ mass values can be excluded, $m_{\etap}^*${}$<$380 MeV with CL$>$99.9\%, regardless of $B^{-1}$. From this preliminary scan, $B^{-1}$=55 MeV can be obtained. In this $B^{-1}$ bin a reasonable mass range, 480$<${}$m_{\etap}^*${}$<$600 MeV corresponds to CL $\geq$ 15\%. This map, however, is not conclusive for the inverse slope parameter $B^{-1}$ in the sense that a subtler analysis, with smaller $B^{-1}$ bins and $m_{\etap}^*$ spectrum-shape dependence is needed to determine an accurate value with proper systematic errors.

The solid line in Fig.\ \ref{f: sim_result} indicates the best choice for the dilepton cocktail, corresponding to $B^{-1}$=55 MeV and $m_{\etap}^*$=520 MeV. Adding \etap in-medium mass modification on top of radial flow and chain decay effects thus further improves the description of the PHENIX dilepton spectrum to a fit quality of $\chi^2/$NDF=15.2/12 and CL =  23\%. Note, however, that Fig.~\ref{f: CL_map} leaves room for further improvements, in particular in the kinematic range where in-medium $\rho$ broadening has been reported~\cite{Arnaldi:2006jq}.

\begin{figure}[h]
 \begin{center} 
   \includegraphics[width=0.45\textwidth]{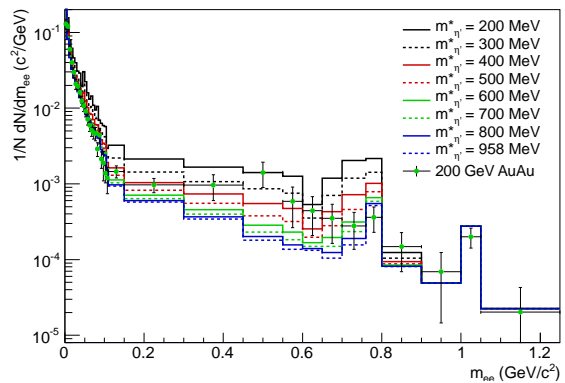} 
\caption{Dilepton cocktail in PHENIX acceptance, for different in-medium $\etap$ mass values. Points are PHENIX data of Ref.~\cite{PPG088}.} \label{f: DiffMasses} 
   \vspace{-10pt}
 \end{center}     
\end{figure}

Fig.\ \ref{f: DiffMasses} shows the dilepton cocktail for different $m_{\etap}^*$ values with a fixed $B^{-1}$=55 MeV. While a suitably chosen $m_\etap^* $ is sufficient to describe the magnitude of the PHENIX low-mass dilepton excess, it leaves room for additional  refinement.

\section{Conclusions}
The current analysis provides a quantitative description of the low-mass enhancement of the PHENIX dilepton spectrum of Ref.~\cite{PPG088}. More details about Section \ref{ssec: radial flow} of the present study can be found in Ref.~\cite{Vargyas, QMposter}. 
As far as we know (see Ref.~\cite{Tserruya:2012jb}), this report is the first that is able to give a statistically acceptable description of the low-mass dilepton enhancement in 200 GeV Au+Au collisions at RHIC. Apparently, radial flow and chain decays have to be taken into account, given that this low-mass dilepton enhancement is in the domain of soft physics with \pt{}$<$2 GeV. In this domain radial flow effects lead to increasing inverse slope parameter with increasing mass. We found that chain decay effects are less important than radial flow if the \etap mass is unchanged. For a dropping \etap mass scenario, however, both effects are important. These effects together provide a data description with CL=1.9\%. 

Considering in-medium \etap mass modification increases the quality of agreement of simulations with data. The best fit values correspond to an in-medium \etap mass of 520 MeV, with a CL=23\%. Extremely low \etap masses below 380 MeV can be excluded at a 99.9\% confidence level. The current results summarize the status of our research until August 2012~\cite{QMposter}. In this work, the error analysis and other systematic studies are still preliminary, and further work is needed to investigate the transverse momentum and centrality dependence of our findings.

\section*{Acknowledgement}

The authors would like to thank I.\ Tserruya, F.\ Geurts and J.\ Sziklai for clarifying discussions.
This research was partially supported by a Charles Simonyi Research Fellowship (T.\ Cs.), by the Hungarian OTKA grant NK 101438, and by the grant CZ.1.07/2.3.00/20.0207 of the ESF programme; Education for Competitiveness Operational Programme.

\end{document}